\documentclass{jaa}
\usepackage{graphicx}

\usepackage{aas_macros} 
\usepackage{natbib}

\begin{document}

\title{Neutron Stars in X-ray Binaries and their Environments}

\author{Biswajit Paul\textsuperscript{1,*}}
\affilOne{\textsuperscript{1}Raman Research Institute, Sadashivanagar, C. V. Raman Avenue, Bangalore 560\,080, India\\}

\twocolumn[{

\maketitle

\corres{bpaul@rri.res.in}

%\msinfo{22 May 2017}{1 January 2018}{1 January 2018}

%%abstract
\begin{abstract}

Neutron stars in X-ray binary systems are fascinating objects that display a wide range of timing and spectral phenomena in the X-rays. Not only parameters of the neutron stars, like magnetic field strength and spin period evolve in their active binary phase, the neutron stars also affect the binary systems and their immediate surroundings in many ways. Here we discuss some aspects of the interactions of the neutron stars with their environments that are revelaed from their X-ray emission. We discuss some recent developments involving the process of accretion onto high magnetic field neutron stars: accretion stream structure and formation, shape of pulse profile and its changes with accretion torque. Various recent studies of reprocessing of X-rays in the accretion disk surface, vertical structures of the accretion disk and wind of companion star are also discussed here. The X-ray pulsars among the binary neutron stars provide excellent handle to make accurate measurement of the orbital parameters and thus also evolution of the binray orbits that take place over time scale of a fraction of a million years to tens of millions of years. The orbital period evolution of X-ray binaries have shown them to be rather complex systems. Orbital evolution of X-ray binaries can also be carried out from timing of the X-ray eclipses and there have been some surprising results in that direction, including orbital period glitches in two X-ray binaries and possible detection of the most massive circum-binary planet around a Low Mass X-ray Binary.

\end{abstract}

%%insert keywords separated by 3 hyphens using \keywords{words}

}]
%%close the twocolumn escape here

%%include \doinum{number}for the DOI number in the header
%%include \volnum{number} for the volume number in the header
%%include \year{yyyy} for  year of publication in the header
%%include \pgrange{num--num} page range of article in the header
%%include \artcitid{num} for the article citation id
%%include \lp to print last page of the article
%%include \setcounter{page}{pagenum} for the exact starting page of the article

%\doinum{12.3456/s78910-011-012-3}
%\artcitid{\#\#\#\#}
%\volnum{123}
%\year{2016}
%\pgrange{23--25}
%\setcounter{page}{23}
%\lp{25}

\section{Introduction}

Neutron stars probably have narrow ranges in their mass and radius (Thorsett \& Chakrabarty 1999), but they are known to have wide range in their other properties like: magnetic field strength, spin period (Konar 2017), rate of change of spin period, age, space velocity, emission mechanism and so on (White, Nagase and Parmar 1995; Psaltis 2006; Kaspi 2010). In some contexts, neutron stars are classified based on absence or presence of binary companion and even based on the nature of their companion star e.g. Low Mass X-ray Binaries (LMXBs) and High Mass X-ray Binaries (HMXBs). The neutron stars in X-ray binary systems not only power themselves by accretion from the companion, they also influence and illuminate their surroundings and the companion stars in many ways, and sometimes help reveal some key information about their surrpundings.

Here we discuss some recent studies on a wide range of topics that could be called as "Neutron stars in X-ray binaries and their environments". Most of these results have been obtained from X-ray timing and X-ray spectroscopic studies from a large number of space observatories, and in some cases by comparing recent data with data acquired several decades ago.

The topics discussed below are disjoint. But we make an effort to bring them up in the order of the distance from the neutron star to the location of the processes involved, starting from the poles of the neutron stars. Many of the topics discussed here, and some of the results reproduced, have been obtained working with my collaborators over a long period.

\section{Disk Magnetosphere Coupling: Accretion Torque and Pulse Profiles} 

The accreting X-ray pulsars provide scope for accurate measurements of their spin period evolution and thus a test of the theory of accretion onto magnetised neutron stars. Two instruments, BATSE onboard CGRO and GBM onboard Fermi have been particularly successful in measuring the spin period evolution of 8-10 bright, persistent, accreting pulsars for about a decade, and for about 30 transient pulsars during their outbursts (Bildsten et al. 1997; Finger et al. 2010 ). In the standard theory of accretion onto magnetised neutron stars, the spin up rate of the neutron star is expected to relate to the X-ray luminosity almost linearly (Ghosh \& Lamb 1979):

\begin{equation}
\dot{\nu} \propto {L_{X}}^{6/7}
\end{equation}
where $\nu$ and L$_{X}$ are the spin frequeny and X-ray luminosity of the neutron star, respectively.

Transient X-ray pulsars, mostly the pulsars in Be X-ray binaries allow excellent tests of the above as these sources traverse upto three orders of magnitude in X-ray luminosity during their outbursts and the dependence of the spin-up rate of the neutron star on the X-ray luminosity is often found to follow the above relation (Ghosh 1996; Bildsten et al. 1997; Sugizaki et al. 2015). These sources are also found to routinely spin-down between outbursts when the X-ray luminosity decreases to quiescence level and the inner radius of any accretion disk becomes larger than the co-rotation radius of the accretion disk for the neutron star's spin period. Centrifugal inhibition of disk accretion may also set in at very low accretion rates.

The persistent sources, on the other hand, do not show any clear dependence between accretion torque and X-ray luminosity. Large changes in accretion torque are often found not to be associated with any significant changes in X-ray luminosity and vice versa (Bildsten et al. 1997; Paul, Rao and Singh 1997; Ikhsanov and Finger 2012; Jenke et al. 2012). This behaviour of the persistent sources, most of which are wind-fed systems with supergiant companion stars is often ascribed to absence of a permanent accretion disk or sometimes due to the presence of counter-rotating accretion disk, a result of accretion from clumps of material from the companion wind that may have net positive or negative angular momentum with respect to the neutron star.

In this context, for a given system, the mass accretion rate is considered to be the only variable factor that determines the accretion torque. How the inner disk magnetically couples to the neutron star has been considered to be identical over a large range of accretion rates. One signature of the nature of coupling of the inner disk to the neutron star poles is in the shape of the accretion column and the resultant X-ray pulse profile. The accretion column or the stream of gas from the inner accretion disk, that is phase locked with the neutron star is known to be narrow. Under some favourable orientation of the system for an observer, the stream or accretion column causes narrow absorption features in the X-ray pulse profile, that has also been spectroscopically confirmed (Galloway et al. 2001; Maitra, Paul and Naik 2012). Any dependence of the X-ray pulse shape with the accretion torque or X-ray luminosity is therefore indicative of changes in the disk-magnetosphere coupling.

\begin{figure}[!t]
\includegraphics[width=1.0\columnwidth, angle=00]{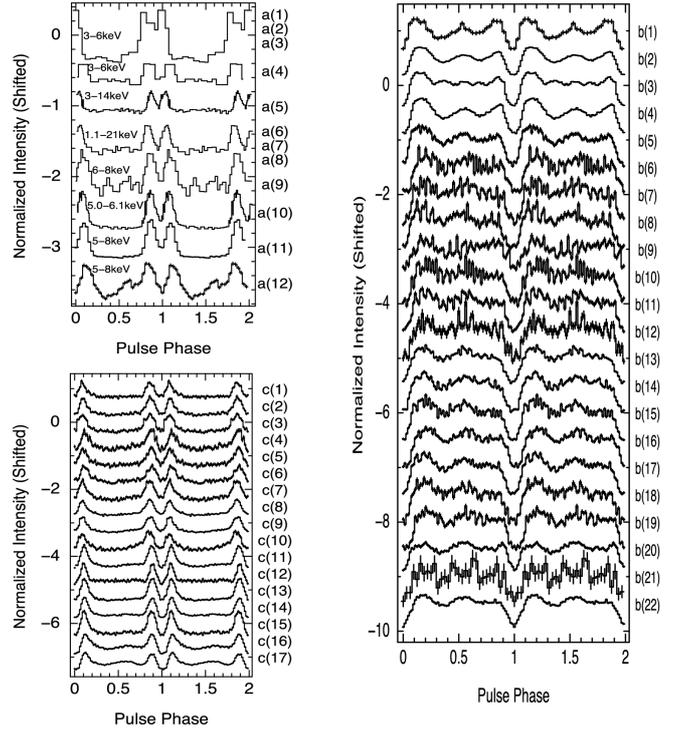}
\caption{The pulse profile of 4U 1626--67 in different accretion torque regimes (from Beri et al. 2014).
Left figures: spin-up phases; right figure: spin-down phase.}\label{figOne}
\end{figure}

Among the persistent accreting X-ray pulsars, 4U 1626--67 is unique in that it shows smooth evolution of pulse period for long periods interspaced with reversal of the accretion torque that happened twice in about last 40 years (Chakrabarty et al., 1997; Jain et al. 2010a). This source therefore offers an excellent test case for changes in the disk magnetophere coupling that can have implication in both accretion torque and X-ray pulse profile. 4U 1626--67 showed monotonous spin-down during 1991-2008 while before and after that period it showed spin-up at a similar spin change timescale in all the three phases. Certain X-ray characteristics of the source are found to be related to the sign of the accretion torque: like strong quasi-periodic oscillations have been seen only in the spin-down era (Kaur et al. 2008) and strong flares have been seen only in the spin-up era (Beri et al. 2014). It was also found that the pulse profile of the source has distinct features that are different in different accreton torque regimes. The profile in the 3-5 keV band, that has been obtained with twelve different X-ray observatories shows a distinct bi-horned shape in the spin-up phases and the shape changes to one with a large dip in the spin-down phase (Beri et al. 2014) as shown in Figure 1.

There are other strong evidences of changes in the accretion column and accretion stream structures with mass accretion rate. During large flares in LMC X-4, its pulse pofile which has a dip during the persistent state changes to a simple sinusoidal shape. The profile also shows a phase shift during the flares compared to the profile just before and after, indicating a change in direction of X-ray beaming (Beri and Paul 2017). A narrow dip in the pulse profile, that is likely produced by absorption of X-rays from the magnetic poles in the accretion stream, disappears during the flare and it takes a few thousands of seconds after the peak of the flare to reappear, indicating the settling timescale of the accretion stream as shown in Figure 2 (Beri and Paul 2017).

\begin{figure}[!t]
\includegraphics[width=.8\columnwidth, angle=00]{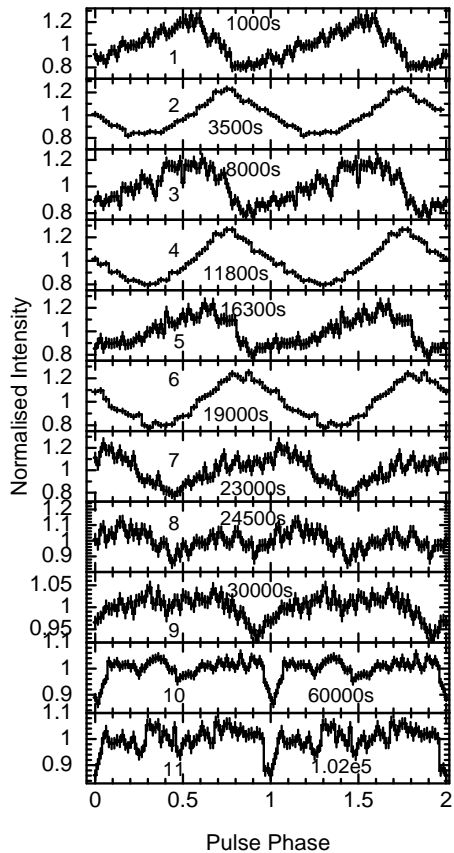}
\caption{Pulse profile evolution of LMC X-4 around several flares is shown here (from Beri and Paul 2017). Time in each panel is with respect to the start of the observation. The profiles in panels 2, 4 and 6 are during the flares. The dip seen in the two bottom panels at phase 1.0 appears about 5000 seconds after the last flare.}\label{figTwo}
\end{figure}

\section{X-ray Reprocessing in Binaries}

The X-rays received from X-ray binary systems are not only those emitted from regions close to the compact stars, but also reprocessed X-rays from various locations. The possible reprocessing sites could be the accretion disk, surface of the companion star, the dense stellar wind, and any intervenning interstellar material. Reprocessed X-rays are very useful to investigate the environments of neutron stars in various ways. Here we give some examples of X-ray reprocessing in different regions and the diagnostics enabled by the same.

\subsection{Reprocessing in Accretion Stream} 

In a certain geometrical configuration of accreting X-ray pulsars, i.e. orientation of the neutron star spin axis with our line of sight,  the accretion stream from the inner accretion disk to the magnetic poles may intercept our view of the polar region at certain spin phases of the neutron star. As a result, the beamed X-rays emitted from the polar region and the accretion column may get absorbed or scattered by the dense accretion stream producing a narrow absorption dip in the pulse profiles. Quite a few X-ray pulsars are now known that show this narrow dip feature and some of them have multiple dips as shown in Figure 3 (Devasia, PhD thesis, 2014). As mentioned before, there is also spectroscopic evidence of increased column density associated with the absorption dips in some pulsars.

\begin{figure*}[!t]
\includegraphics[width=2.0\columnwidth, angle=00]{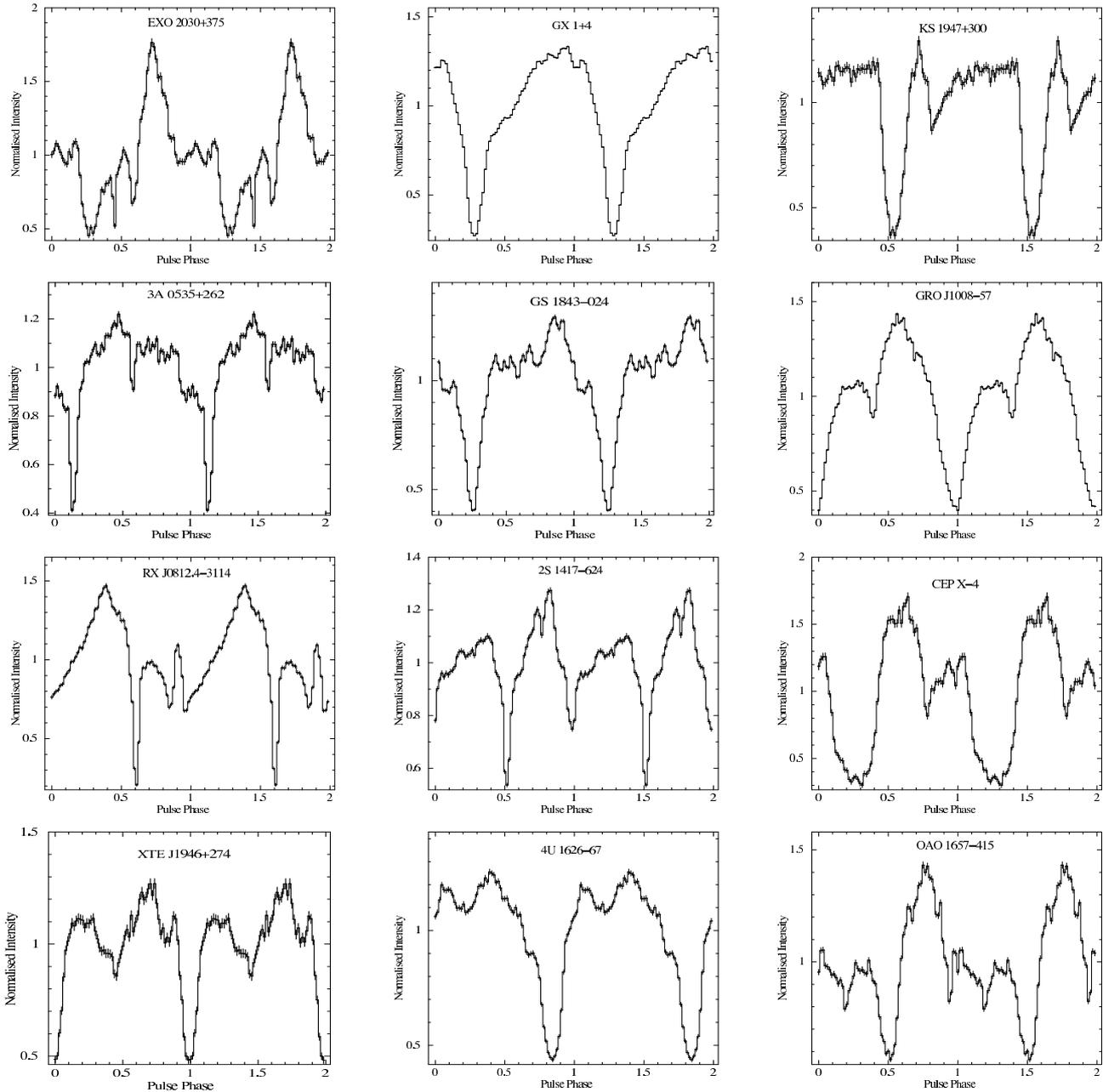}
\caption{A collection of pulse profiles from various accretion powered pulsars showing narrow dip features (Devasia, PhD thesis 2014).}\label{figThree}
\end{figure*}

\subsection{Reprocessing in Accretion Disk}

The accretion disk is likely to occupy a substantial solid angle as seen from the central X-ray emitting region and can thus reprocess the X-rays. Reprocessing of X-rays from the accretion disk can produce both emission lines and continuum emission. In low magnetic field neutron stars, in which the accretion disk can extend to very close to the neutron star, line emission shows signatures of gravitational redshift (Bhattacharyya and Strohmayer 2007; Cackett et al. 2010) and for lines produced at a larger distacnce, velocity broadening due to Keplerian motion of materail in the accretion disk has been observed (4U 1626--67, Schulz et al. 2001). In the case of accretion powered high magnetic field pulsars, the accretion disk has an inner radius of a few hundred km. While the disk temperature does not reach the X-ray regime by viscous dissipation, X-ray heating achieves it. For an X-ray luminosity of $\sim10^{37}$ erg s$^{-1}$ of the central source, and an inner radius of several hundred km, the disk reprocesses the central X-rays and emits blackbody radiation at a characteristic temperature of 100-200 eV (Paul et al 2002). A different origin of this soft spectral component in accreting X-ray pulsars, which otherwise show a hard power-law type X-ray spectrum has been found to modulate differently with respect to the hard X-rays as shown in Figure 4 (Paul et al. 2002). However, this feature is detectable only in the few X-ray pulsars that are away from the Galactic plane and do not suffer from absorption of the soft X-rays by a large column density. viz. Her X-1, SMC X-1 and LMC X-4. Interestingly, all these three sources also have a superorbital intensity modulation at timescales of several times the orbital period, presumably caused by a precessing warped accretion disk. In accordance with this scenario, the phase differene of the soft X-rays with respect to the power-law continuum has also been found to vary with the superorbital phase (Neilsen, Hickox and Vrtilek 2004).

\begin{figure}[!t]
\includegraphics[width=1.0\columnwidth, angle=-90]{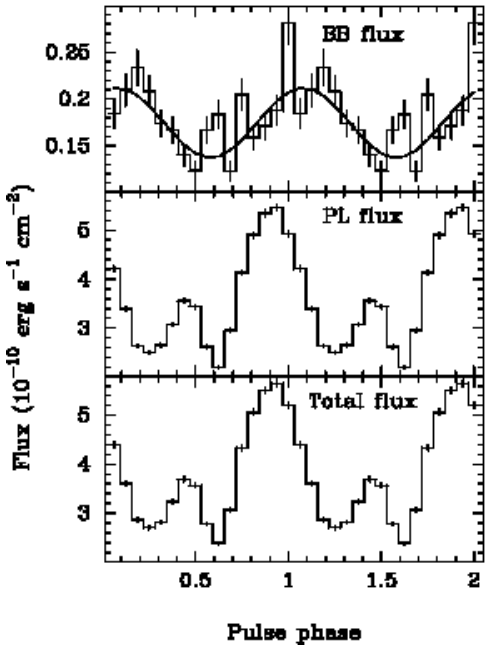}
\caption{Pulse profiles of the blackbody component, power-law component and total emission of SMC X-1 are shown here (from Paul et al. 2002). The blackbody component clearly has a different pulse profile and phase with respect to the rest of the X-ray emission and shows a different origin.}\label{figFour}
\end{figure}

Thermal reprocessing of the X-rays by the outer accretion disk is in the UV-Optical regime which has been investigated for several decades (two extensive reviews are van Paradijs \& McClintock 1995 and Charles \& Coe 2006). An interesting manifestation of the optical reprocessing is seen in the LMXBs. During the thermonuclar X-ray bursts in the LMXBs, the X-ray luminosity increases by a factor of $\sim10$ in about a second, sometimes reaching Eddington luminosity and then fades with a timescale of $\sim10$ seconds (Galloway et al. 2008). The X-ray burst with this sharp timing feature, is reprocessed in the accretion disk and often increases the total optical emission from the binary by a factor of a few (Pedersen et al. 1982 ; van Paradijs \& McClintock 1995; Hynes et al. 2006). Any delay and smearing of the optical emission with respect to the X-rays from the central source carries information about the light travel time from the neutron star to the reprocessing region or time across the reprocessing region. The extended nature of the reprocesing region is likely to cause smearing of the UV-optical burst. Study of the X-ray to optical reprocessing of the thermonuclear bursts at different orbital phases of LMXBs can provide valuable information about the binary system, akin to imaging of such compact systems (Hynes et al. 2006, Paul et al. 2012). Moreover, simultaneous X-ray and multi-wavelength UV-Optical observations can give useful information about the reprocessing phenomena itself (Hynes et al. 2006).

The emission lines of various ionisation species of oxygen and neon that are produced by reprocessing of X-rays from the accretion disk of 4U 1626--67 show pulse phase dependence of the flux as shown in Figure 5, which is rare among X-ray pulsars (Beri, Paul \& Dewangan  2015). In this source, the pulse phase dependence indicates presence of vertical structures in the accretion disk with azimuthal asymmetry.

\begin{figure}[!t]
\includegraphics[width=.95\columnwidth, angle=00]{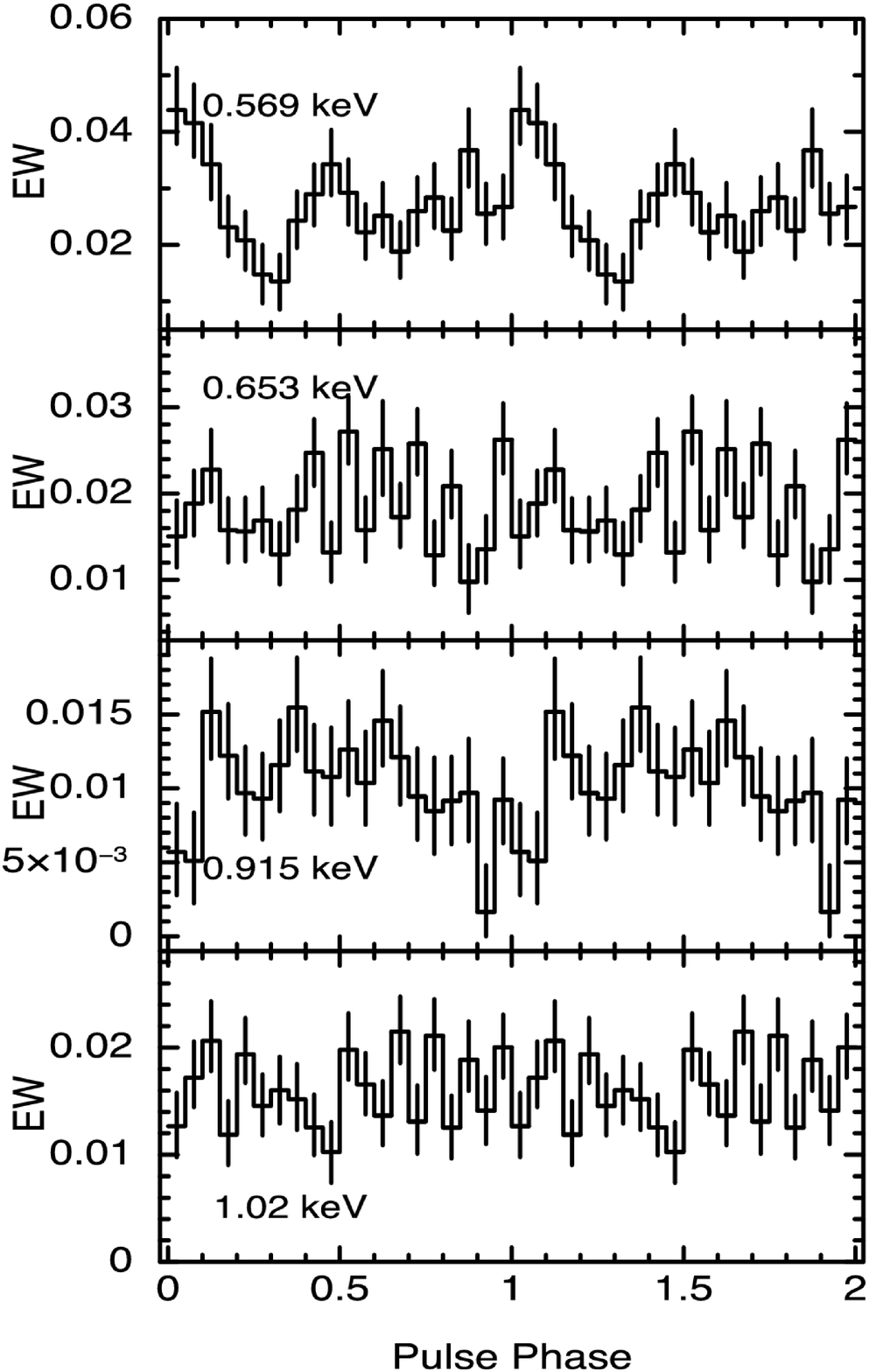}
\caption{The flux of various emission lines in the X-ray spectrum of 4U 1626--67 are shown here as a function of the pulse phase (Beri et al. 2015). A strong pulse phase dependence can be clearly seen in the top panel.}\label{figFive}
\end{figure}

\subsection{Reprocessing from the Surface of the Companion Star}

LMXBs are more suitable than the HMXBs for detection of X-rays reprocessed from the surface of the companion star. Perhaps the best example of this phenomenon is reprocessing of pulsed X-ray emission from the LMXB pulsar 4U 1626--67 (Middleditch et al. 1981; Raman et al. 2016). In this source, with a nearly face on orientation of the orbital plane with respect to our line of sight, the reprocessed optical pulses are detected with the same pulse period as the X-ray pulses and a second set of pulses is detected with a period that is slightly offset. The second set of pulses has a smaller pulse amplitude and is due to reprocessing from the surface of the companion star and hence with a period that is a beat frequency between the X-ray pulsar's spin period and orbital period of the binary.

\begin{figure}[!t]
\includegraphics[width=.95\columnwidth, angle=-90]{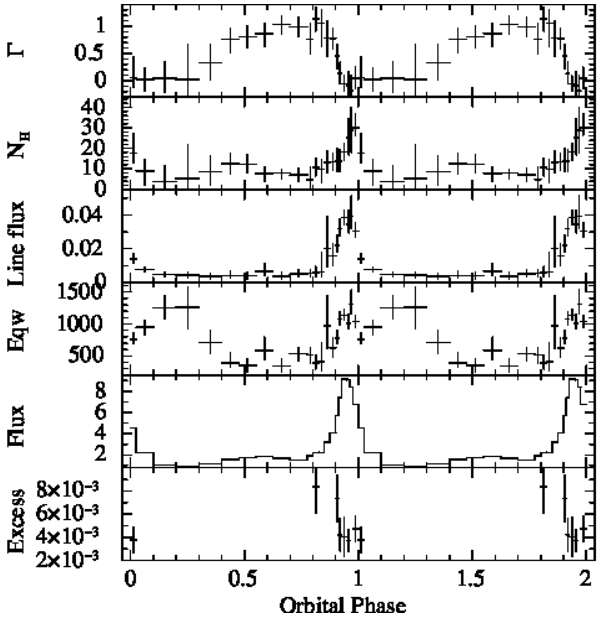}
\caption{Variation of the X-ray spectral parameters of GX 301--2 are shown here with orbital phase. At the orbital phase 0.2, the iron line equivalent width shown in the fourth panel is very large while the hydrogen column density of absorption, shown in the second panel is low (from Islam and Paul 2014). Unit of the two parameters are eV and 10$^{22}$ cm$^{-2}$, respectively.}\label{figSix}
\end{figure}

\subsection{Reprocessing from the Stellar Wind}

The massive companion stars in the HMXBs lose mass at the rate of about 10$^{-6}$ $M_{\odot}$ yr$^{-1}$ via stelalr wind with a terminal velocity of $\sim1000$ km s$^{-1}$. The HMXB wind is therefore another reprocessing agent of central X-ray sources and two prominent signatures of the reprocessing in stellar wind are iron K$_\alpha$ line emission (Gim{\'e}nez-Garc{\'{\i}}a et al. 2015) and absorption of the soft X-rays. The magnitude of the reprocessing varies largely from system to system: the iron line equivalent width can be negligible, $\sim20$ eV (SMC X-1: Naik and Paul 2004b) to as large as 1.5 keV (GX 301--2: Islam and Paul 2014 ) and the column density can be less than 10$^{21}$ cm$^{-2}$ (LMC X-4: Naik and Paul 2004a) to several times 10$^{24}$ cm$^{-2}$ (GX 301--2: Islam and Paul 2014). Predictably, both of these parameters also show strong variation over the orbital phase, especially in binaries with large eccentricity (GX 301--2). In certain orbital phases of GX 301--2, the neutron star goes through an extremely dense component of the stellar wind, and the iron K$_\alpha$ line photons suffer Compton scattering in the same medium producing a prominent Compton recoil feature which can in turn be used to determine the iron abundance in the medium and also its temperature (Watanabe et al. 2003).

The wind of the companion star, even in the supergiant companions, is not homogeneous and also not isotropic. One manifestation of the neutron star passing through/near the inhomogeneity/clumpiness of the stellar wind is that the mass accretion rate changes at short timescales and the view of the reprocessing medium (i.e, a clump close to the neutron star) also changes at short timescales. This results in changes in the X-ray flux, column density of absorbing material, and iron line equivalent width at short time scales. Prime examples of the same are IGR J17544--2619 (Rampy et al. 2009), IGR J18410-0535 (Bozzo et al. 2011) and 
 OAO 1657--415 (Pradhan et al. 2015).

\begin{figure}[!t]
\includegraphics[width=.65\columnwidth, angle=-90]{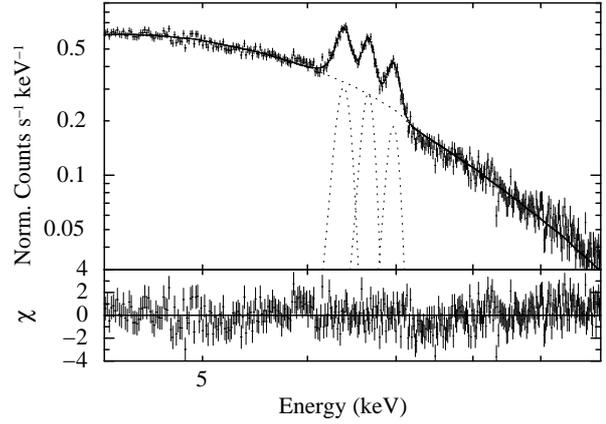}
\caption{The X-ray spectrum of Cen X-3 taken during an eclipse egress showing three different iron line components from neutral, helium-like and hydrogen-like iron atoms (from Naik and Paul 2012).}\label{figSeven}
\end{figure}

The orbital dependence of the absorption column density and iron line equivalent width can be a very useful tool to understand the mass accretion process in some HMXBs. An HMXB that shows a large intensity modulation with orbital phase is GX 301--2. The binary orbit of GX 301--2 is significantly eccentric ($e=0.47$: Sato et al. 1986, but a radial distance dependence of the mass accretion rate, or orbital phase dependent X-ray absorption does not explain the strong orbital intensity modulation of this source. What is most peculiar about GX 301--2 is that the X-ray intensity peaks about 2 days before the periastron passage of the neutron star in its orbit around the companion. An inclined outflowing disk from the companion star (Pravdo and Ghosh 2001) and an accretion stream flowing out from the companion star (Leahy and Kostka 2008) have been proposed to explain the large variation in X-ray luminosity and mass accretion rate. However, both of these models require the iron line equivalent width and the absorption column density to vary in a certain pattern over the orbital period. These parameters, measured with MAXI-GSC, instead show a different pattern of variation with orbital phase (Islam and Paul 2014). An intriguing feature observed at phase 0.2 after the periastron passage is that the iron line equivalent width is very large (greater than 1.0 keV) while the absorption column density is moderate (less than 10$^{23}$ cm$^{-2}$)) as shown in Figure 6. This observation can only be explained with a dense medium behind the neutron star at this orbital phase, which is different from the two models above that had been proposed for this source.

\begin{figure}[!t]
\includegraphics[width=.65\columnwidth, angle=-90]{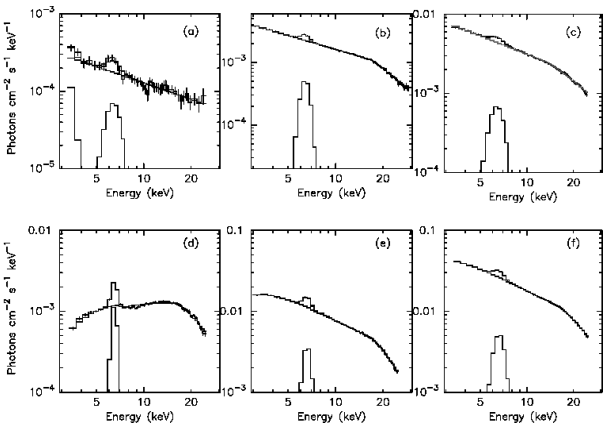}
\caption{The X-ray spectra of LMC X-4 (top) and Her X-1 (bottom) are shown here
from left to right during their low, medium and high intensity states, respectively, of their superorbital periods. The iron emission line is more dominant in the low states (from Naik and Paul 2003). }\label{figEight}
\end{figure}

\subsection{Reprocessing and X-ray Eclipses}

The eclipsing X-ray binaries, about ten of each type LMXBs and HMXBs, are excellent objects to study the reprocessing of X-rays in the stellar wind or in the outer accretion disks. The reprocessed emission is usually a small fraction of the total emission received from these objects and in normal conditions, the direct component dominates. But during the X-ray eclipses, in absence of the source emission, all that is detected is reprocessed. In HMXBs, the iron line and other emission lines are produced in the stellar wind, and the size of the reprocessing region is several times the size of the companion star. During the eclipses of the X-ray source, the comtinuum X-ray emission which also goes through scattering in the ionised wind is suppressed by a factor of 10-100, but the line fluxes are reduced by a factor of only a few. Therefore, during the eclipses, the emission lines dominate the spectrum (van der Meer et al. 2005). In almost all HMXBs, the dominating component of the iron line is from neutral or near neutral species producing a 6.4 keV line. One excpetion is Cen X-3, which also has highly ionised species: helium-like and hydrogen-like iron producing three iron lines at 6.4 keV, 6.7 keV and 6.95 keV (Figure 7). The evolution of the relative strengths of the iron lines during the eclipse egresss of Cen X-3 is enchanting, showing that the highly ionised species are further away from the neutron star than the neutral iron atoms (Naik and Paul 2012).

Another interesting scenario in which reprocessed X-ray emission shows predominance over the source continuum emission by masking the compact object is in the sources with superorbital variations like Her X-1 and LMC X-4. The low states of their superorbital period are known to be due to obscuration of the compact objects by precessing warped accretion disks. Similar to what happens during the X-ray eclipses, the equivalent width of the iron emission line becomes much larger, in excess of 1.0 keV during the low states of the superorbital period as shown in Figure 8 (Naik and Paul 2003).

\section{Orbital Evolution of X-ray Binaries}

The X-ray binaries evolve and evolution of their orbital periods has been measured with a varying degree of accuracy in different kinds of systems with obsevations carried out over long periods of a few years to upto a few decades. The most accurate of the methods is by pulse timing when the compact object is a pulsar. The LMXBs have sharp eclipse transitions that take place over $\sim$10 seconds that allows accruate measurement of the mid-eclipse times. This in turn allows measurement of orbital period evolution of LMXBs which has a time scale of several tens of million years, to be carried out within a few years. One LMXB, 4U 1822--37 has a large accretion disk corona and being an extended source, the X-ray emission from 4U 1822--37 is not completely eclipsed. Even its partial eclipses have been useful for the same purpose (Jain et al. 2010b). A few binary systems like Cyg X-3 and and 4U 1820--30 which do not have eclipses, but a stable orbital modulation pattern; the pattern itself has been used as a time marker and the same has allowed measurement of the period derivatives (Chou and Grindlay 2001; Singh et al. 2002). In the absence of any useful time marker in the X-ray band, orbital evolution measurements of a few black hole X-ray binaries have been carried out successfully albeit with lesser accuracy, using the Doppler shift of the companion star spectra
(Gonz{\'a}lez Hern{\'a}ndez et al. 2017).

\begin{figure}[!t]
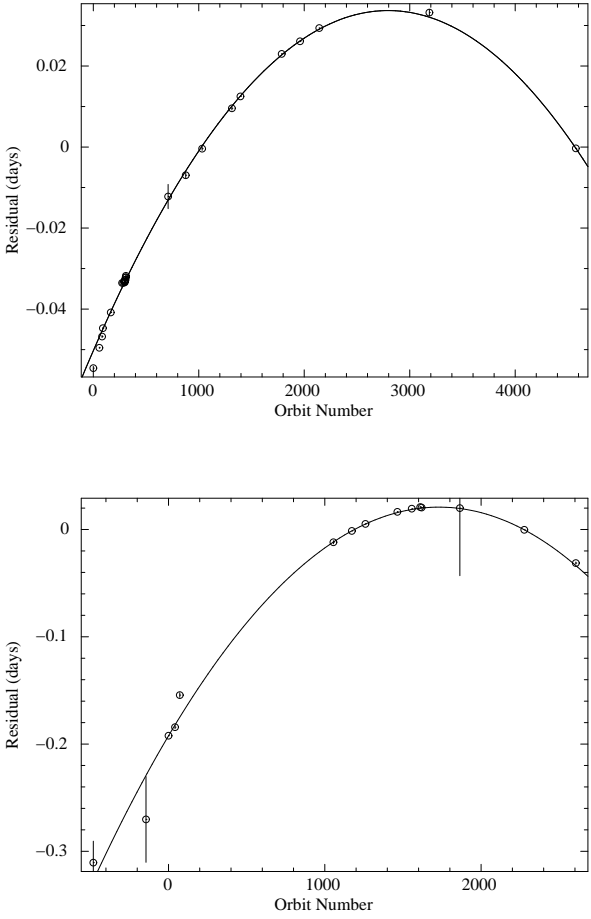

\includegraphics[width=.65\columnwidth, angle=-90]{Cen_X-3.eps}
\vskip 1cm
\includegraphics[width=.65\columnwidth, angle=-90]{SMC_X-1.eps}
\caption{Orbital evolution of two HMXBs, Cen X-3 (top) and SMC X-1 (bottom) are shown here. The measured mid-eclipse times are plotted after subtracting a linear term (from Raichur and Paul 2010a).}\label{figNine}
\end{figure}

\subsection{Evolution of High Mass X-ray Binaries}

Pulse timing measurements over a few decades with a range of timing instruments have resulted in determination of the orbital period evolution in several HMXBs like Cen X-3 and SMC X-1 (Raichur and Paul 2010a), LMC X-4 (Naik and Paul 2004a), OAO 1657--415 (Jenke et al. 2012) and 4U 1538--52 (Mukherjee et al. 2006). Multiple measurements of the mean longitude $T_{\pi/2}$ with pulse timing show a non-linear term in it as shown in Figure 9, which is a measure of the rate of orbital period evolution. Here we note that the eclipse timing itself can be used to meausre orbital evolution in quite a few HMXBs, and the difference between the orbital period of an eccentric binary measured with X-ray eclipses and the same measured with pulse timing can be used to determine the apsidal advance rate, if any, of an HMXB (Falanga et al. 2015). Apsidal advance has also been detected in the transient X-ray binary 4U 0115+63 using pulse timing (Raichur and Paul 2010b). In the absence of pulsations, eclipse measurements have been used in 4U 1700--37 (Islam and Paul 2016), the rate of evolution of which was earlier overestimated as the effect of intrinsic rapid intensity variation of the source during the eclipse transitions was not taken into account.
With the exception of Cyg X-3, the orbital periods of the HMXBs are found to decay. The first four HMXBs mentioned above show an orbital evolution time scale of less than a million years, which is much faster than the wind mass loss time scale of greater than 10$^{7}$ years. This implies that the orbital evolution is driven primarily by tidal interaction between the two stars. However, in spite of having similar binary parameters (companion star mass and orbital period), two of the HMXBs, 4U 1538--52 and 4U 1700--37 show a much smaller rate of period decay, which is surprising.

The decaying orbital period of HMXBs is intersting in itself. A high rate of orbital decay does indicate towards a large no of such systems thay may have evolved away in the past. The known number of supergiant HMXBs in the galaxy, and their current rate of orbital decay can be used to estimate the no of such systems that have evolved away in the past and obtain independent estimates of their descendents in the Milky Way, and by extension, to other star forming galaxies. Vast majority of the HMXBs with a neutron star are expected to have the neutron star spiraling into its companion during a Common-Envelope phase after onset of Roche-lobe overflow in them (Bhattacharya and van den Heuvel 1991; Taam and Sandquist 2000). However, as discussed in the previous paragraph, in spite of having similar binary parameters, some HMXBs show significantly different, and lower, orbital evolution rate. Owing to these differences, if a fraction of the HMXBs survive as X-ray binaries  and eventually produce NS-NS and NS-BH binaries, then HMXB evolution has a wider implication. The double compact binaries will finally merge by gravitationl wave radiation producing a fast gamma-ray burst and a burst of graviational waves. The number of potential such sources of gravitationl waves are usually estimated from the observed number of double neutron star systems, which has limited statistics.

\begin{table*}[ht]
\tabularfont
\caption{List of LMXBs with negative orbital period derivative, arranged in order of the orbital decay timescale.}\label{tableOne}
\begin{tabular}{lcccccc}
\topline
Name & Compact Star & $P_{orb}$ & $\dot{P}_{orb}$ & $P_{orb}\over{-2\dot{P}_{orb}}$ & Reference\\
& Type & (hr) & (s s$^{-1}$) & (10$^{6}$yr) & \\\midline
Nova Muscae 1991 & BH & 10.4 & --6.6 x 10$^{-10}$ & 0.9 & Gonz{\'a}lez Hern{\'a}ndez et al. 2017\\
XTE J1118+480 & BH & 4.1 & --6.0 x 10$^{-11}$ & 3.9 & Gonz{\'a}lez Hern{\'a}ndez et al. 2014\\
4U 1820--30 & NS & 0.19 & --1.1 x 10$^{-12}$ & 9.9 & Peuten et al. 2014 \\
AX J1745.6--2901 & NS & 8.4 & --4.0 x 10$^{-11}$ & 12.0 & Ponti et al. 2017 \\
A0620--00 & BH & 7.8 & --1.9 x 10$^{-11}$ & 23.4 & Gonz{\'a}lez Hern{\'a}ndez et al. 2014\\
MXB 1658--298 & NS & 7.12 & --1.2 x 10$^{-11}$ & 33.9 & Jain et al. 2017 \\
Her X-1 & NS & 40.8 &  --4.8 x 10$^{-11}$ & 48.5 & Staubert et al. 2009 \\
\hline
\end{tabular}
\end{table*}

\subsection{Evolution of Low Mass X-ray Binaries}

Similar to the HMXBs, the orbital period evolution of LMXBs have also been measured with the techniques of pulsar timing in some sources and with eclipse timing or orbital intensity profile in some other sources. In some of the sources, the existing data is consistent with the orbital period evolving in a secular manner while at least in two LMXBs, the data suggests sudden changes in orbital period, called orbital period glitches. Among the LMXBs in which pulsar timing has been used, Her X-1 and 4U 1822--37 are high magnetic field pulsars, while all the other sources are Accreting Milli-second X-ray Pulsars (AMXPs). Perhaps with the exception of IGR J0029+5934 (Patruno 2017) all the classical LMXBs show negative orbital evolution or orbital period glitches. On the other hand, AMXPs may show positive orbital period evolution.

\subsubsection{Secular Evolution:}

Two Accreting Millisecond X-ray Pulsars (AMXPs), SAX J1808.4-−3658 (Jain et al. 2008, Hartman et al. 2009, Patruno et al. 2016) and SAX J1748.9−-2021 (Sanna et al. 2016) show secular and positive orbital period evolution with a timescale $\left(P_{orb}\over{2\dot{P}_{orb}}\right)$ of 4-65 million yeas, considerably smaller than the mass transfer timescales of these transient sources. 4U 1822--37 also shows a positive orbital evolution, measured from its partial eclise with a timescale of 2 million years (Jain et al. 2010b). The fast positive evoltuion compared to what would be expected due to a conservative mass transfer in these systems could be due to non-conservative mass transfer in which a signficant amount of mass is lost from the system, either as disk-wind or as outflow from the inner Lagrangian point (Patruno et al. 2016). High orbital evolution rates have also been measured in Black Widow and Redback binary pulsar systems (see Patruno et al. 2016 for a list) but we don't discuss them here.

Other than Her X-1 (Staubert et al. 2009), two LMXBs AX J1745.6--2901 (Ponti et al. 2017) and MXB 1658--298 (Jain et al. 2017) show negative orbital period change rate. Interestingly, three black hole X-ray binaries XTE J1118+480, A0620--00 and Nova Muscae 1991 also show negative orbital period evolution at high rates (Gonz{\'a}lez Hern{\'a}ndez et al. 2014, 2017). A list of LMXBs that show negative orbital period evolution is given in Table 1.

\begin{figure}[!t]
\includegraphics[width=0.7\columnwidth, angle=-90]{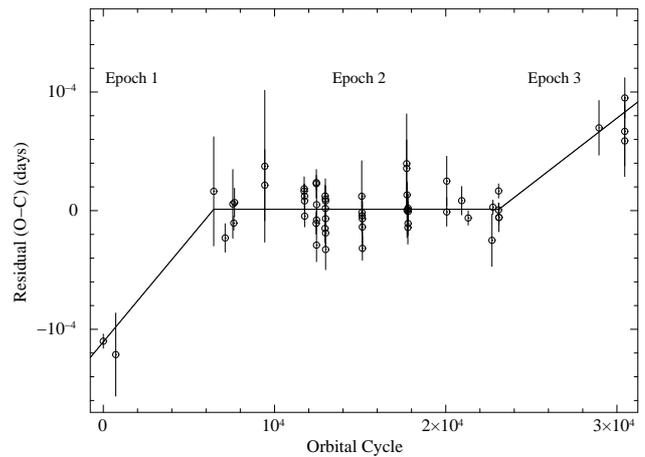}
\caption{The mid-eclipse times of XTE 1710--281 are shown here after subtrcting a linear term (from Jain et al. 2011).}\label{figTen}
\end{figure}

\subsubsection{Orbital Period Glitches:}

Two of the eclipsing LMXBs, EXO 0748--676 (Wolff et al. 2009), and XTE J1710--281 (Jain \& Paul 2011; Figure 10), show orbital period glitches in which the orbital period is nearly constant over a certain period of time and shows changes from epoch to epoch by a few milliseconds or $\left( {\Delta{P_{orb}}}\over{P_{orb}}\right)$ of about 10$^{-7}$. In both of these systems, the glitches have been measured using their mid-eclipse times. The timescale over which this change takes place could be very short in EXO 0748--676 (Wolff et al. 2009), while in XTE J1710--281, it could be very short to upto a few years.

The reasons behind the orbital period glitches are not well understood. This and the rapid negative orbital evolutions mentioned for some LMXBs above have also been ascribed to the Applegate mechanism (Applegate and Patterson 1987) which is due a variable gravitational quadrapole of a companion star with high magnetic field. It is important to note that in either of the systems EXO 0748--676 and XTE J1710--281, no significant flux change was associated with the episodes of the orbital period glitches.

\begin{figure}[!t]
\includegraphics[width=1.0\columnwidth, angle=00]{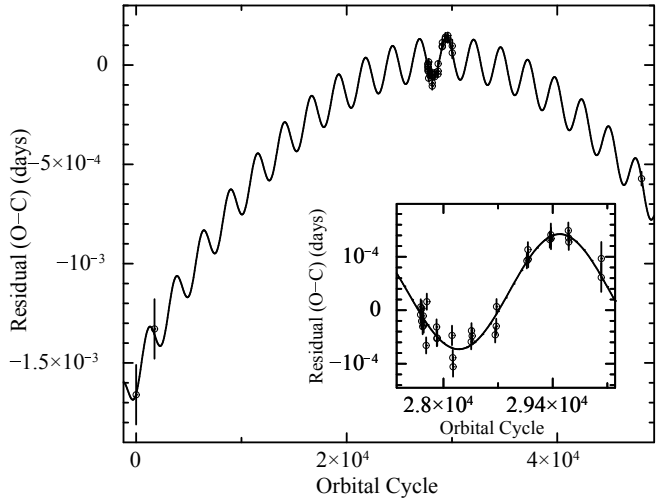}
\caption{The mid-eclipse times of MXB 1658--298 are shown here after subtracting a linear term. The line represents the best fit function to the data with a quadratic term and a sinusoidal term (taken from Jain et al. 2017. A part of the data is shown expanded, separately in the inset.)}\label{figEleven}
\end{figure}

\subsubsection{A Massive Circum-Binary Planet around MXB 1658--298:}

In addition to a secular change in the mid-eclipse times due to an orbital decay with a timescale $\left(P_{orb}\over{2\dot{P}_{orb}}\right)$ of $\sim$32 million years, the mid-eclipse times of MXB 1658--298 shows a periodic variation with an amplitude of ~9 seconds and a period of about two years (Figure 11). A simple interpretation of this is to be due to the presence of a massive circum-binary planet of mass about 23 Jupiter masses at an orbital radius of about 1.5 AU (Jain et al. 2017). The periodic delay of eclipse time has subsequently also been detected in the most recent ourburst of the source (Iaria et al. 2017), which is a prediction for the presence of a circum-binary planet.

If true, it is the only known planet around an X-ray binary system and has been detected thanks to periodic delays of X-ray eclipse timings. It is also the most massive among about 20 circum-binary planets known and the host system is the most compact one among all known binary stellar systems around which planets have been found. The host system, being an LMXB is also likely to be very old. This planet is therefore likely to be an inportant input in the study of circum-binary planet formation, migration, and survival.

Along with the presence of a massive planet around MXB 1658--298, a rapid orbital decay of its binary orbit is interesting.
Several LMXBs with neutron star or black hole compact objects show a rapid orbital period decay (Table 1). It is therefore interesting to investigate if angular momentum exchange of the X-ray binary with the planet play a role in the orbital decay of MXB 1658--298 and whether such objects are present around other LMXBs as well.

\section{Conclusion}

Neutron stars are fascinating objects and are cosmic laboratories to investigate some extremes of the physical universe like high density nuclear matter, extremely high magnetic field strength, different manifestations of strong gravity etc.
In this article, we have discussed some aspects of astrophysical studies that are enabled by neutron stars in a binary setting.

\section*{Acknowledgement}

Author thanks two reveiwers who provided important feedback on the issues of orbital evolution of X-ray binaries. The author is grateful to his collaborators with whom most of the original works rediscussed in this article were carried out and also thanks the editors for their initiave to bring out this special issue.

%-------------------------------------------------------------------------------

\end{document}